\documentclass{svjour2} 
\smartqed 
\usepackage{graphicx}
\usepackage[T1]{fontenc}
\usepackage{amssymb, latexsym, amsopn}

\begin{document}

\title{The Szekeres Swiss Cheese model and the CMB observations}

\author{Krzysztof Bolejko}

\institute{The University of Melbourne, Melbourne VIC 3010, Australia, and \\
Nicolaus Copernicus Astronomical Center,
	   Bartycka 18, 00-716 Warsaw, Poland \\
	  \email{bolejko@camk.edu.pl}}

\date{Received: date / Accepted: date}

\maketitle

\begin{abstract}
This paper presents the application of the Szekeres Swiss Cheese model to 
the analysis of  observations  of the cosmic microwave background (CMB) radiation. 
The impact of inhomogeneous matter distribution 
on the CMB observations is in most cases studied
within the linear perturbations of the Friedmann model.
However, since the density contrast and the 
Weyl curvature within the cosmic structures are large, 
this issue is worth studying using another approach.
The Szekeres model is an inhomogeneous, non-symmetrical and exact solution of 
the Einstein equations.
In this model, light propagation and matter evolution can be  exactly calculated, without such approximations   as small amplitude of the density contrast. This 
allows to examine in more realistic manner the contribution of the light propagation effect to the measured CMB temperature fluctuations.

The results of such analysis show that small-scale, non-linear inhomogeneities induce, 
via Rees-Sciama effect, temperature fluctuations of amplitude $10^{-7} - 10^{-5}$
on angular scale $\vartheta< 0.24^{\circ}$ ($\ell > 750$). 
This is still much smaller 
than the measured temperature fluctuations on this angular scale.
However, local and uncompensated inhomogeneities can induce temperature fluctuations of amplitude as large as $10^{-3}$, and thus can be responsible the
low multipoles anomalies observed in the angular CMB power spectrum.
\end{abstract}


\section{Introduction}\label{intro}
The Universe, as it is observed, is very inhomogeneous. Among structures observed in the Universe are clusters and superclusters of galaxies as well as large cosmic voids. This inhomogeneous matter distribution affects light propagation and hence
astronomical observations. 
The study of light propagation effects is considerably important when
analysing the CMB observations. This is because the last
scattering surface is the most remote region
which is  observable using electromagnetic radiation.
In standard approach the CMB temperature fluctuations are analysed 
by solving the Boltzmann equation
within linear perturbations around the 
homogeneous and isotropic Friedmann--Lema\^itre--Robertson--Walker (FLRW) model \cite{SZ96,SSWZ03}\footnote{This approach is implemented in such
codes like  CMBFAST (http://www.cfa.harvard.edu/$\sim$mzaldarr/CMBFAST/cmbfast.html), CAMB (http://www.camb.info/),
or CMBEASY (http://www.cmbeasy.org/).}.
The application of the FLRW model as a background model
results in a remarkably good fit to the CMB data
\cite{H08}.  However, the assumption of homogeneity, which is also consistent
with other types of cosmological observations
is not a direct consequence of them \cite{E08}.
It is often said that such theorems 
like the Ehlers-Geren-Sachs theorem \cite{EGS} and the `almost EGS theorem' \cite{SME}
justify the application of the FLRW models. These theorems state
 that if anisotropies in the cosmic microwave background radiation are small for all fundamental observers then locally the Universe is almost spatially homogeneous and isotropic. 
However, as shown in \cite{NUWL99},
it is possible that the CMB temperature fluctuations are small
but the Weyl curvature is large. In such a case the geometry of the Universe is far 
from the Robertson--Walker geometry and the applicability of FLRW models
is not justified.
Moreover, the applicability of the linear approach can
be questionable since the density contrast within cosmic structures
is much larger than unity. Therefore, there is a need for application of exact and inhomogeneous models to the study of the light propagation and its impact on the CMB temperature fluctuations. This issue has been extensively studied within spherically symmetric models ---
within the thin shell approximation \cite{TV87,IS06,IS07}
and  within the Lema\^itre--Tolman model \cite{P92,AFMS93,SAF93,AFS94,FSA94,RRS06,MN08}. 
However, most of the cosmic structures are not
spherically symmetric, and thus the study of light propagation in
non-spherical models is essential. One of the suitable models for this purpose is the Szekeres model.
 The Szekeres model has no symmetry, allows to study a nonlinear evolution and does not require 
small Weyl curvature.
Therefore, this paper aims to study the CMB temperature fluctuations 
in the Swiss Cheese Szekeres model.

The structure of this paper is as follows.
Sec. \ref{lighp} presents basic formulae describing
light propagation; in Sec. \ref{szekmdl} the Szekeres model 
is presented; Sec. \ref{setup} presents the construction of models;
Sec. \ref{results} presents the findings.

\section{Light propagation}\label{lighp}

Light propagates along null geodesics. If  $k^{\alpha}$ is a vector tangent to a
null geodesic, then
\begin{equation}
k_{ \alpha} k^{\alpha} = 0, \qquad k_{ \alpha ; \beta} k^{\beta} = 0.
\label{ngeo}
\end{equation}
As light propagates, the frequency of photons changes. 
The ratio of the frequency of a photon at the
emission event to the measured frequency defines the redshift
\begin{equation}
\frac{\nu_e}{\nu_o} := 1+z.
\end{equation}
Since photon's energy, as measured by an observer with the 4-velocity $u^{\alpha}$, is proportional to $k^{\alpha} u_{\alpha}$, thus the redshift obeys the following relation

\begin{equation}
1 + z = \frac{\left( k^{\alpha} u_{\alpha} \right)_e}{ \left( k^{\alpha}
u_{\alpha} \right)_o}, \label{rdfdef}
\end{equation}
where the subscripts $_e$ and $_o$ refer to instants of emission and
observation respectively.
Assuming that the black body spectrum is conserved,
the temperature must be proportional to $1+z$
\begin{equation}
\frac{T_e}{T_o} = 1 + z. \label{tmpscl}
\end{equation}
Then, from eq. (\ref{tmpscl}), the temperature fluctuations measured by co-moving
observer are:
\begin{equation}
\left( \frac{\Delta T}{T} \right)_o = \frac{T_e/(1+z) - \bar{T}_e / (1 +
\bar{z})}{\bar{T}_e/(1+\bar{z})}, \label{tmpfl}
\end{equation}
where quantities with bars $~ \bar{}~$ refer to average quantities.
If the temperature at the emission is $T_e = \bar{T}_e + \Delta T_e$, then
\begin{equation}
\left( \frac{\Delta T}{T} \right)_o = \frac{\bar{z} - z}{1+z} + \left( \frac{\Delta T}{\bar{T}} \right)_e
\frac{1+\bar{z}}{1+z}. \label{dtpt}
\end{equation}
As can be seen from the above formula, the observed temperature fluctuations on
the CMB sky are caused by the light propagation effects and by the
temperature fluctuations at the decoupling instant.

\section{The Szekeres model}\label{szekmdl}

\subsection{The metric of the Szekeres model}

For our purpose it is convenient to use a coordinate system different from
that in which Szekeres \cite{S75} originally found his solution. The metric is of the following form \cite{C96}

\begin{equation}
{\rm d} s^2 =  c^2 {\rm d}t^2 - \frac{(\Phi,_r - \Phi E,_r/E)^2}
{(\varepsilon - k)} {\rm d} r^2 - \Phi^2 \frac{({\rm d} p^2 + {\rm d} q^2)}{E^2}, \label{ds2}
 \end{equation}
where $\Phi$ is a function of $t$ and $r$, $\varepsilon = \pm1,0$ and $k = k(r)
\leq \varepsilon$ is an arbitrary function of $r$.
The function $E$ is given by
 \begin{equation}
E(r,p,q) = \frac{1}{2S}(p^2 + q^2) - \frac{P}{S} p - \frac{Q}{S} q + C ,
\label{edef}
 \end{equation}
where the functions $S = S(r)$, $P = P(r)$, $Q = Q(r)$, and $C = C(r)$ satisfy
the relation
 \begin{equation}
C = \frac{P^2}{2S} + \frac{Q^2}{2S} + \frac{S}{2} \varepsilon,~~~~~~~~~
\varepsilon = 0, \pm 1.
 \end{equation}

Originally, Szekeres considered only a case of $p=0=\Lambda$.
This result was generalised by Szafron \cite{Szf77} to the case of uniform pressure, $p=p(t)$. A spacial case of this solution, the cosmological constant, was in detailed discussed 
by Barrow and Stein-Schabes \cite{BSS84}.

The $\varepsilon = -1$ case is often called the  
quasihyperbolic Szekeres model (for a detailed discussion on the 
quasihyperbolic Szekeres models see \cite{HK08}),
$\varepsilon = 0$ quasiplane  (for details see \cite{HK08,K08}),
and $\varepsilon = 1$ quasispherical (for details see \cite{HK02}).
Although it is possible to have within one model quasispherical  and 
quasihyperbolic regions separated by the quasiplane regions \cite{HK08},
only the quasispherial case will be considered here.

In the quasispherial case surfaces of constant $t$ and $r$ are spheres.
The transformation from ($p,q$) coordinates into ($\vartheta$, $\varphi$)
coordinates is \cite{HK02}

\begin{eqnarray}
&& p = S \cot \frac{\vartheta}{2} \cos \varphi + P, \nonumber \\
&& q = S \cot \frac{\vartheta}{2} \sin \varphi + Q.
\label{stp} 
\end{eqnarray}

\subsection{The Einstein equations}

Applying metric (\ref{ds2}) to the Einstein equations, and assuming the
energy momentum tensor for a dust, the Einstein equations reduce
 to the following two

\begin{equation}
\frac{1}{c^2} \Phi,_t^2 = \frac{2M}{\Phi} - k + \frac{1}{3} \Lambda
\Phi^2, \label{vel}
\end{equation}

\begin{equation}
4 \pi \frac{G}{c^2} \rho =  \frac{M,_r - 3 M E,_r/E}{\Phi^2 ( \Phi,_r - \Phi E,_r/E)}. \label{rho}
\end{equation}
where $\rho$ is matter energy density, $M(r)$ is another function of radial coordinate.
In a Newtonian limit $M c^2/G$ is equal to the mass inside the shell of radial coordinate
$r$.  However, it is not an integrated rest mass but  active gravitational mass that generates a gravitational field.

Eq. (\ref{vel}) can be integrated

\begin{equation}
\int\limits_0^{\Phi}\frac{{\rm d}
\tilde{\Phi}}{\sqrt{\frac{2M}{\tilde{\Phi}} - k + \frac{1}{3} \Lambda
\tilde{\Phi}^2}} = c (t- t_B), \label{cal}
\end{equation}
where $t_B(r)$ is an arbitrary function of
$r$. This means that the big bang is not a single event as in the FLRW
models but occurs at different times for different distances from the origin.

As can be seen the Szekeres model is specified by 6 functions. However, by a
choice of the appropriate coordinates, the number of independent functions can be reduced
to 5.

\subsection{General properties and the Friedmann limit}

The vorticity within the Szekeres model is zero. In addition
the acceleration vanishes, $u^{\alpha}{};_{\beta} u^{\beta} = 0$. The shear
tensor is
\begin{equation}\label{shear}
\sigma^{\alpha}{}_{\beta} = \frac{1}{3} \left( \frac{\Phi,_{tr} - \Phi,_t
\Phi,_r/ \Phi} {\Phi,_r - \Phi E,_r/E} \right) {\rm diag}
(0,2,-1,-1).
\end{equation}
The scalar of expansion is
\begin{equation}\label{expansion}
\theta =  u^{\alpha}{}_{;\alpha} = \frac{\Phi,_{tr} + 2 \Phi,_t \Phi,_r/ \Phi -
3 \Phi,_t E,_r/E}{ \Phi,_r - \Phi E,_r/E}.
\end{equation}

The Weyl curvature decomposed into its electric and magnetic part is of the
following form
\begin{eqnarray}\label{Weyl}
&& E^{\alpha}{}_{\beta} = C^{\alpha}{}_{\gamma \beta \delta} u^{\gamma} u^{\delta} =
\frac{M(3 \Phi,_r - \Phi M,_r/M)}{3 \Phi^3 ( \Phi,_r - \Phi E,_r / E)}
{\rm diag} (0,2,-1,-1), \nonumber \\
&& H_{\alpha \beta} = \frac{1}{2} \sqrt{-g} \rho_{\alpha \gamma \mu \nu} C^{\mu
\nu}{}_{\beta \delta} u^{\gamma} u^{\delta} = 0.
\end{eqnarray}
Finally, the 4D and 3D Ricci scalars are
\begin{eqnarray}\label{Ricci}
&& ^4\mathcal{R} = - \kappa \rho c^2 - 4 \Lambda , \nonumber \\
&& ^{3}\mathcal{R} = 2 \frac{k}{\Phi^2} \left( \frac{ \Phi k,_r/k - 2 \Phi
E,_r/E}{ \Phi,_r - \Phi E,_r/E} + 1 \right).
\end{eqnarray}

In the Friedmann limit, $\Phi \rightarrow r a_F$,  $k \rightarrow k_F r^2$
and $M \rightarrow M_F r^3$
where $a_F$ is the Friedmann
scale factor, $k_F$ is the curvature index and is a constant, and $M_F$ is another constant.
As can be seen in Friedmann limit:

\begin{eqnarray}
& \theta & \rightarrow 3 \frac{a_{F},_t}{a_F}, \nonumber \\
& \sigma^{\alpha}{}_{\beta} & \rightarrow 0, \nonumber \\
& E^{\alpha}{}_{\beta} & \rightarrow 0, \nonumber \\ 
& ^4\mathcal{R} & \rightarrow - 6  \frac{M_F}{a_F^3}  - 4 \Lambda, \nonumber \\ 
& ^{3}\mathcal{R} & \rightarrow 6 \frac{k_F}{a_F^2}.
\label{flim}
\end{eqnarray}

\subsection{Null geodesic equations}\label{light}

The geodesic equations

\begin{equation}
\frac{{\rm d}^2 x^{\alpha}}{{\rm d} s^2} + \Gamma^{\alpha}_{\beta \gamma}
\frac{{\rm d} x^{\beta}}{{\rm d} s} \frac{{\rm d} x^{\gamma}}{{\rm d} s} = 0.
\label{geoeq}
\end{equation}
in the quasispherical Szekeres model, are of the following form

$\alpha = 0$:
\begin{eqnarray}
 && \frac{{\rm d}^2 t}{{\rm d} s^2} + \frac{{\Phi,_{tr}} - {\Phi,_t} {E},_r/E}{1 - k} (\Phi,_r - \Phi E,_r/E)  \left( \frac{{\rm
d} r}{{\rm d} s} \right)^2 \nonumber \\
&& +  \frac{\Phi {\Phi,_t}}{E^2} \left[ \left(
\frac{{\rm d} p}{{\rm d} s} \right)^2
 + \left( \frac{{\rm d} q}{{\rm d} s} \right)^2 \right] = 0,
\label{lp0}
\end{eqnarray}

$\alpha = 1$:
\begin{eqnarray}
&& \frac {{\rm d}^2 r} {{\rm d} s^2} + 2 \frac {{\Phi,_{tr}} - {\Phi,_t}{E},_r/E}{\Phi,_r - \Phi E,_r/E} \frac{{\rm d} t}{{\rm d} s}
\frac{{\rm d} r}{{\rm d} s} 
- \frac{\Phi}{E^2} \frac{1-k}{\Phi,_r - \Phi E,_r/E}
\left[ \left( \frac{{\rm d} p}{{\rm d} s} \right)^2 + \left( \frac{{\rm d}
q}{{\rm d} s} \right)^2 \right]
\nonumber \\
&& + \left[ \frac{\Phi,_{rr} - \Phi,_r E,_r/E - \Phi {E},_{rr}/E + \Phi (E,_r/E)^2}{\Phi,_r - \Phi {E},_r/E} + \frac{1}{2}\frac{k,_r}{1-k} \right] \left( \frac{{\rm d}
r}{{\rm d} s} \right)^2 \nonumber \\
&& + 2 \frac{\Phi}{E^2} \frac{E,_r  E,_p - E
 E,_{pr}}{\Phi,_r - \Phi E,_r/E} \frac{{\rm d}
r}{{\rm d} s} \frac{{\rm d} p}{{\rm d} s} + 2 \frac{\Phi}{E^2}
\frac{(E,_r E,_q - E  {E},_{qr})}{\Phi,_r - \Phi E,_r/E } \frac{{\rm d} r}{{\rm d} s}
\frac{{\rm d} q}{{\rm d} s}   = 0,
 \label{lp1}
\end{eqnarray}

$\alpha = 2$:

\begin{eqnarray}
&& \frac{{\rm d}^2 p}{{\rm d} s^2} + 2 \frac{{\Phi,_t}}{\Phi} \frac{{\rm d}
t}{{\rm d} s}  \frac{{\rm d} p}{{\rm d} s} - \left[ \frac{{\Phi},_r
- {\Phi} E,_r/E}{\Phi(1 - k)} (E,_r  E,_p - {E}  E,_{pr}) \right] \left(\frac{{\rm d} r}{{\rm d} s} \right)^2 + \frac{ E,_p}{E} \left( \frac{{\rm d} q}{{\rm d} s} \right)^2
\nonumber \\
&& + 2 \left(  \frac{\Phi,_r}{\Phi} - \frac{E,_r}{E} \right)
\frac{{\rm d} r}{{\rm d} s}  \frac{{\rm d} p}{{\rm d} s} -
 \frac{ E,_p}{E} \left( \frac{{\rm d} p}{{\rm d} s}
 \right)^2 - 2 \frac{E,_q}{E}
\frac{{\rm d} p}{{\rm d} s}  \frac{{\rm d} q}{{\rm d} s}  =
0, 
\label{lp2}
\end{eqnarray}

$\alpha = 3$:
\begin{eqnarray}
&& \frac{{\rm d}^2 q}{{\rm d} s^2} + 2 \frac{{\Phi,_t}}{\Phi} \frac{{\rm d}
t}{{\rm d} s}  \frac{{\rm d} q}{{\rm d} s} - \left[ \frac{\Phi,_r
- {\Phi} E,_r/E}{\Phi (1 - k)} (E,_r  E,_q - {E}  E,_{qr}) \right] \left( \frac{{\rm d} r}{{\rm d} s} \right)^2 - \frac{E,_q}{E} \left( \frac{{\rm d} q}{{\rm d} s} \right)^2
\nonumber \\
&& + 2 \left( \frac{\Phi,_r}{\Phi} - \frac{E,_r}{E} \right)
\frac{{\rm d} r}{{\rm d} s}  \frac{{\rm d} q}{{\rm d} s} + \frac{ {E},_q}{E} \left( \frac{{\rm d} p}{{\rm d} s} \right)^2 - 2 \frac{{E},_p}{E} \frac{{\rm d} p}{{\rm d} s} \frac{{\rm d} q}{{\rm d} s}  = 0. 
\label{lp3}
\end{eqnarray}

Equations (\ref{lp0}) -- (\ref{lp3}) are quite complicated.
However, if two coordinates could be constant on a geodesic, then
we could  impose ${\rm d} p = {\rm d} q = 0$
on a general solution of (\ref{ds2}) to get
\begin{equation}
 \frac{{\rm d} t}{{\rm d} r} = \pm  \frac{\Phi,_r - \Phi E,_r/E}{\sqrt{1 -
 k}},
\label{snge}
\end{equation}
where $+$ is for outwards directed geodesics and $-$ for inwards directed geodesics.
In such a case the redshift formula (\ref{rdfdef}) 
 would reduce to the much simpler form (see Appendix \ref{app1} for derivation)
\begin{equation}
\ln (1+z) =  \pm \int\limits_{r_e}^{r_o} {\rm d} r
\frac{ \Phi,_{tr} - \Phi,_t E,_r/E}{\sqrt{1 - k}}.
\label{srf}
\end{equation}
Thus, if such fixed-direction geodesics exist, then the study of light
propagation in the Szekeres model could be significantly simplified. Instead of
solving eqs. (\ref{lp0}) -- (\ref{lp3})
only eq. (\ref{snge}) would have to be solved
to find a null geodesic and only eq. (\ref{srf}) to find the
redshift.
However, in general ${\rm d}^2 p/{\rm d} s^2 \neq 0$ and
${\rm d}^2q/ {\rm d}s^2 \neq 0$, i.e. the condition ${\rm d} p = {\rm d} q = 0$
cannot hold along the null geodesic.
As follows from (\ref{lp0}) -- (\ref{lp3}) if initially $k^p= k^q =0$, then the
coordinates $p$ and $q$ will remain constant only if along the whole geodesic
\begin{equation}
\Phi,_r = \Phi \frac{E,_r}{E}, \label{cd1}
\end{equation}
or
\begin{equation}
E E,_{pr} = E,_p E,_{r}~~~{\rm and}~~~E E,_{qr} =
E,_q E,_{r}. \label{cd2}
\end{equation}
The relation (\ref{cd1}) holds only at a shell crossing singularity which must be eliminated
 in a physically acceptable model. 
Apart from the spherical symmetry (i.e. the Lema\^itre--Tolman model)
the relations (\ref{cd2}) hold only in the axially symmetric case
\cite{ND07}. It should be noted that not every Szekeres model is 
axially symmetric. Moreover, apart from the spherical symmetry
there is only one such geodesic,
which propagates along the symmetry axis. Thus, such a geodesic will be referred to as the axial geodesics. 

\section{The Swiss Cheese model}\label{setup}

\subsection{Arrangement of the Swiss Cheese model}

\begin{figure*}
\includegraphics[scale=0.56]{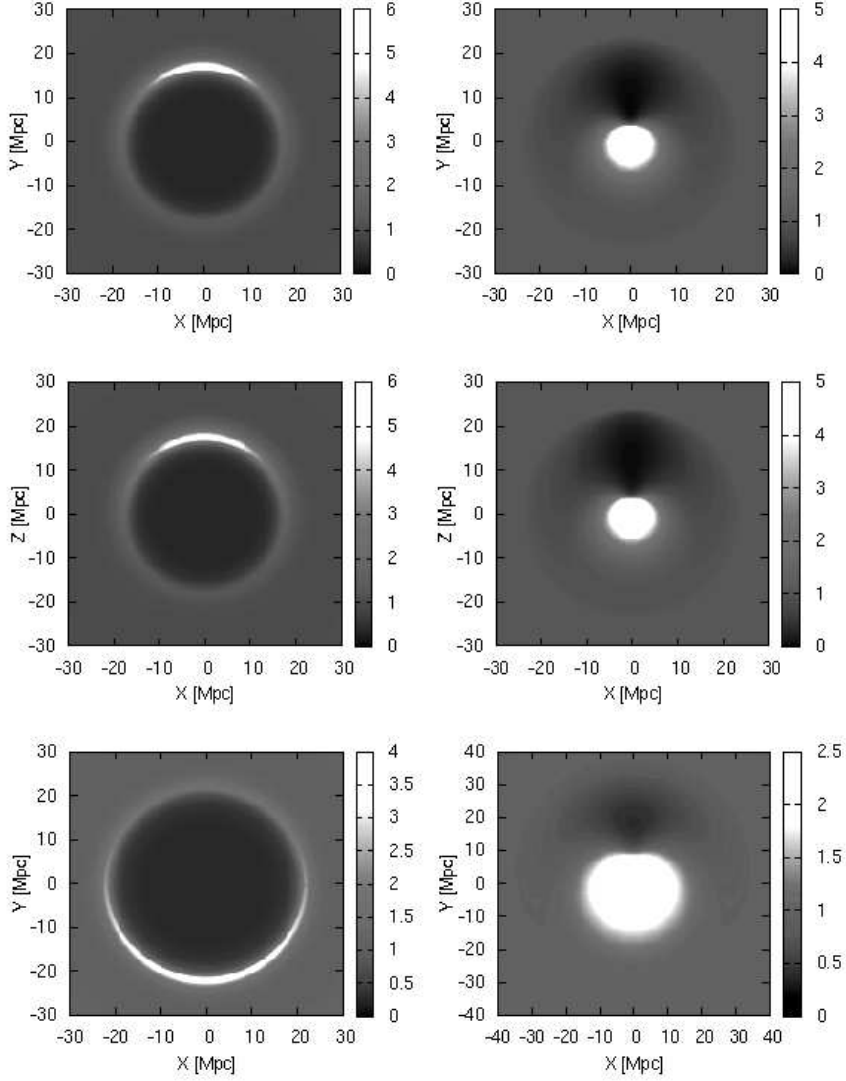}
\caption{The density distribution, $\rho / \rho_b$, at the current instant for 
regions A (upper left), B (upper right), C (middle left), D (middle right),
E (lower left) and F (lower right). 
In regions A, B, E and F the dipole axis is aligned with the Y axis.
For these regions the density distribution is presented for the surface $Z=0$.
In regions C and D the dipole is aligned with the Z axis. 
For these regions the density distribution is presented for the surface $Y=0$.
Coordinates $X,Y,Z$ are defined as $X= \Phi \cos \varphi \sin \vartheta, 
Y =\Phi \sin \varphi \sin \vartheta, Z =\Phi \cos \vartheta$ [where 
$\vartheta$ and $\varphi$ are given by (\ref{stp})].
The exact forms of functions used to specify
these models are included in Appendix \ref{app2}.}
\label{fig1}
\end{figure*}

\begin{figure}
\includegraphics[scale=0.52]{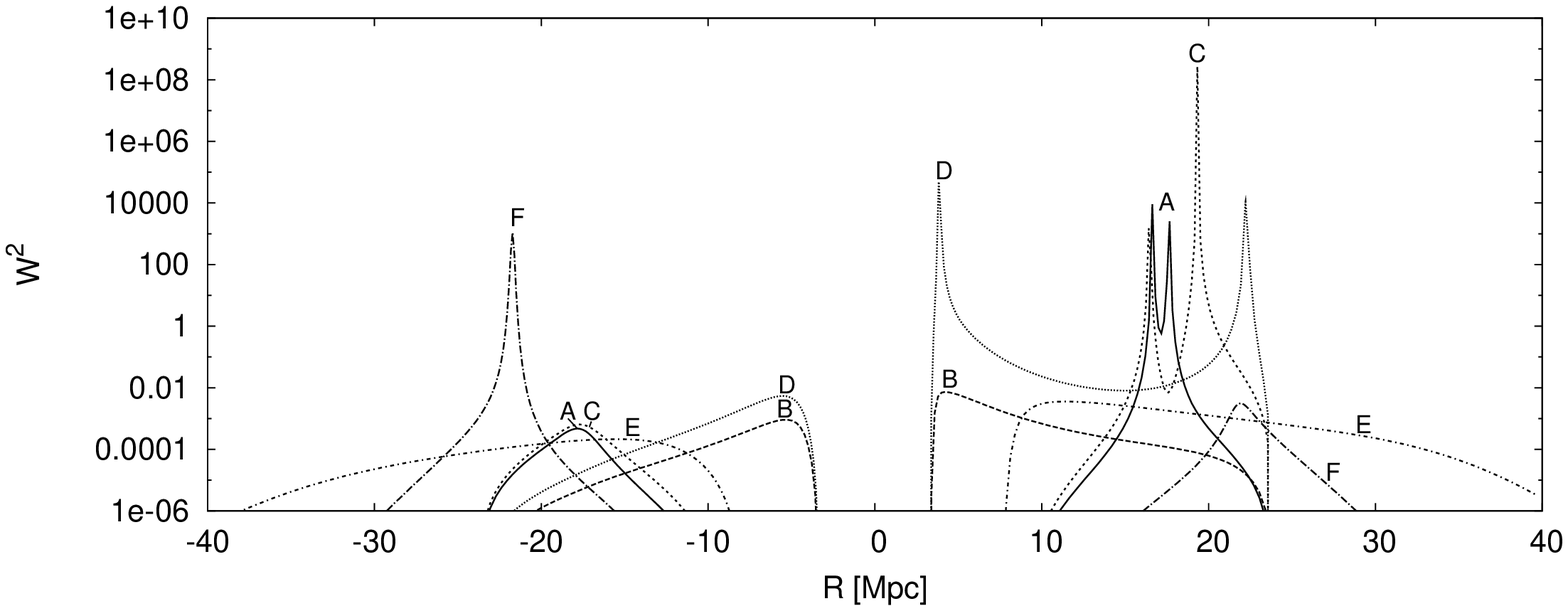}
\caption{The Weyl curvature scalar ${\mathcal W}^2$ [eq. (\ref{Weyl})] evaluated at the current instant,
along the dipole axis, R, in Szekeres regions A--F (for regions A, B, D, and E, R =Y, for regions C and D, R=X).
The 3D shape of these profiles is very similar to the shape 
of the density distribution -- see Fig. \ref{fig1}.
Since the FLRW models are conformally flat, the Szekeres regions
are in some parts far from being even a close (linear) approximation of the FLRW model.}
\label{fig2}
\end{figure}

The Swiss Cheese models which are employed in this paper 
are constructed from six different building blocks -- regions A-F (holes) --
which are matched with the Friedmann background (cheese).
These Szekeres patches are placed so that
their boundaries touch wherever a light ray exits one inhomogeneous patch.
Thus
the ray immediately enters another Szekeres inhomogeneity and does not spend any time in
the Friedmann background. 
Using different sequences of regions A -- F five models are constructed, .
The density distribution at the current instant within each of these regions is
presented in Fig.\ \ref{fig1}. 
The exact forms of functions used to specify
the Szekeres model in each of these regions are
presented in Appendix \ref{app2}.
As can be seen from the detailed specification (Appendix \ref{app2})
the functions defining regions A--D
become for the radial cooridate\footnote{The
radial coordinate in this paper is defined by the value of $\Phi$
at the last scattering instant: $r = \Phi_{LS}$ -- see Appendix \ref{app2}.
Thus $r=24$ kpc corresponds to the current distance of
approximetaly 26 Mpc -- cf. Fig. \ref{fig1}.} $r \geqslant 24$ kpc
of exactly the same form as the form in the Friedmann background 
[compare the form of the functions in Appendix \ref{app2} with the 
form of functions in the Friedmann limit, eqs. (\ref{flim})].
 Regions E and F tend
exponentially to the Friedmann models. However, as seen from their specification
or from Fig.\ \ref{fig2}, at the distance $r \approx 30$ kpc and
$r \approx 40$ kpc, respectively, regions E and F become almost Friedmann.
Figure \ref{fig2} presents the curvature scalar, ${\mathcal W}^2$, which is
defined as
\begin{equation}
\mathcal{W}^2 = \frac{E_{\alpha \beta} E^{\alpha \beta}}{6 H^4}, \label{7.4}
\end{equation}
where $E_{\alpha \beta}$ is the electric part of the Weyl tensor (\ref{Weyl})
and $H = (1/3) \theta$ is the Hubble parameter (\ref{expansion}).
As can be seen, in some regions ${\mathcal W}^2 \gg 1$. This feature, apart from
nonlinear evolution and non-symmetrical shape, makes the application of the
Szekeres model more realistic.

\subsection{Junction conditions}

When constructing a Swiss Cheese model, we need to satisfy the junction
conditions for matching the particular inhomogeneous patches to the Friedmann
background, and also assure the continuity of the null geodesics.  The standard
junction conditions are that the 3D-metric of the surface and its extrinsic
curvature, the first and second fundamental forms, must be continuous.

For matching a Szekeres patch to a Friedmann background across a comoving
spherical surface, $r =$ constant, the conditions are: that the mass inside the
junction surface in the Szekeres patch is equal to the mass that would be inside
that surface in the homogeneous background;
that the spatial curvature  at the junction surface is the same in both
the Szekeres and Friedmann models, which implies that $k_{SZ} = k_F r^2$
and $(k_{SZ}),_r = 2 k_F r$; finally that the bang time and also $\Lambda$ must be continuous across the junction.
The mass $M$ and the curvature function $k$
are matched by the construction --- see Appendix \ref{app2}.
The value of the cosmological constant is the same in all regions,
and the value of the bang time function is fixed by (\ref{cal}),
and at the junction is equal to $t_B=0$.
It might be surprising that a non-symmetrical model
like the Szekeres  can be joined with the symmetric FLRW model, but
there are other examples of such junctions. For example Bonnor demonstrated that the
Szekeres model can be matched to the Schwarzschild solution \cite{B76}.

The junction of null geodesics requires the continuity of
all components of the null vector.
However, let us notice that when one Szekeres sphere is matched to another Szekeres
sphere it can be rotated around the normal direction. Thus, we only need to match
up the time component, $k^0$ and  the tangential component. The 
tangential component is defined as
 \begin{equation}
   k^T = \frac{\Phi}{E} \sqrt{ (k^p)^2 + (k^q)^2}.
\label{ktdf} 
\end{equation}
The radial component is then given by the null condition, $k_{\alpha} k^{\alpha}
=0$. 

\subsection{Description of models}

Five different Szekeres Swiss Cheeses models are considered here:

\begin{enumerate}

\item Model 1

Model 1 is constructed from alternately matching regions A and B (A + B + A + B
...) into the Friedmann background. When a light ray  exits one Szekeres region,
it immediately enters another inhomogeneous patch. Each time the $(p,q)$
position of the point of entry is randomly selected. In addition $k^p$ and $k^q$
are quasi-randomly selected, i.e
\[ (k^q)^2 = \gamma \left( k^T \frac{E}{\Phi} \right)^2,
 \quad (k^p)^2 = (1 - \gamma) \left( k^T \frac{E}{\Phi} \right)^2, \]
where $\gamma$ is a random value in the range $0 \leqslant \gamma \leqslant 1$.
The radial coordinate of the matching point is $r_j = 24$ kpc -- the point where
the Szekeres region becomes Friedmann.

\item Model 2

This model is constructed from alternating regions C and D, but only axial null
geodesics are considered, i.e.\  $k^p=0$ and $k^q = 0$, $p=q=0$. The radial
component of the matching point is again $r_j = 24$ kpc.  

\item Model 3

The next model consists of regions E and F placed alternately. Null vector
components $k^p$ and $k^q$ are chosen in such a way that
$10^{-8} \leqslant k^p \leqslant 10^{-4}$ and $10^{-8} \leqslant k^q \leqslant 10^{-4}$,
but are otherwise random. As can be noted, this is not in accordance with
condition (\ref{ktdf}). In order to maintain the continuity of the tangential
component of the null vector the next Szekeres patch must by reoriented with
respect to the preceding  patch. This however leads to an overlapping of these
two Szekeres regions. Although, at the junction point ($r_j = 40$ kpc for region
E, and $r_j = 50$ kpc for region F), these two regions are almost Friedmann,
still this is not a perfect matching. We proceed with this type of imperfect
matching to study how randomly chosen values of the tangential component (hence
more randomised light propagation through a Szekeres patch) influences the final
results.

\item Model 4

Model 4 is constructed using only C regions, with $r_j = 24$ kpc, and only axial
geodesics are considered, i.e.\ $k^p=k^q = p = q = 0$.

\item Model 5

The last model is also axially symmetric, $k^p=k^q = p = q = 0$, $r_j = 24$ kpc,
but uses only D regions.

\end{enumerate}

\section{Results}\label{results}

\subsection{The Rees-Sciama effect}

\begin{figure}
\includegraphics[scale=0.355, angle=270]{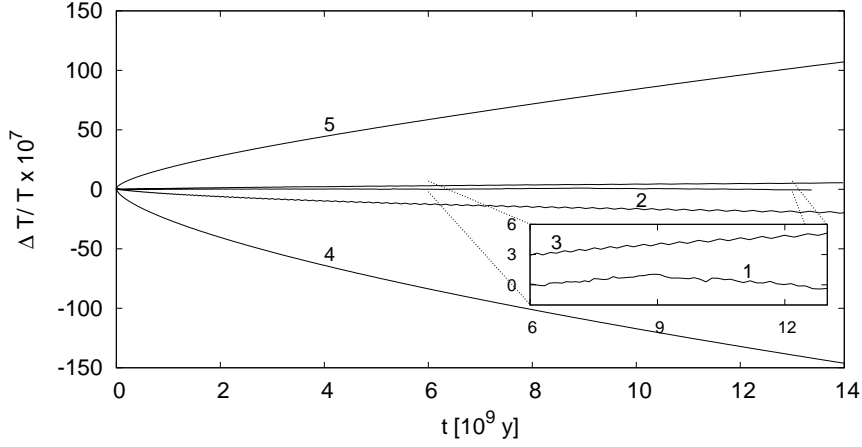}  
\caption{The temperature fluctuations caused by light propagation effects in models 1-5.
In models 1-3 light propagates alternatively though underdense and overdense regions.
In model 4 light propagates only though regions of $\delta M < 0, \delta k >0$,
and in model 5 only through regions of $\delta M > 0, \delta k < 0$ (see 
Sec. \ref{setup} for a detail description on how these models were constructed).}
\label{fig3}
\end{figure}

\begin{figure}
\includegraphics[scale=0.235, angle=270]{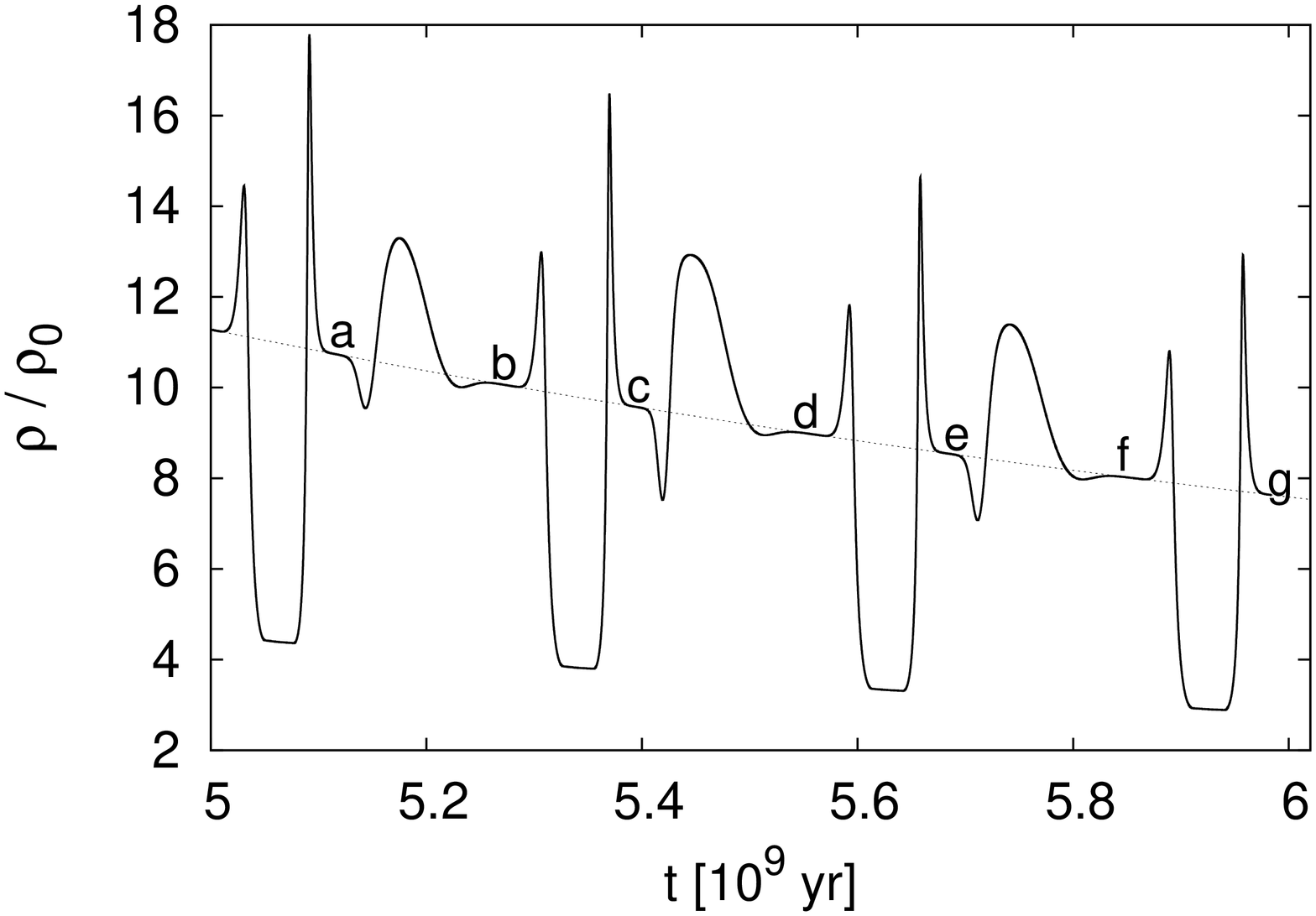}
\includegraphics[scale=0.235, angle=270]{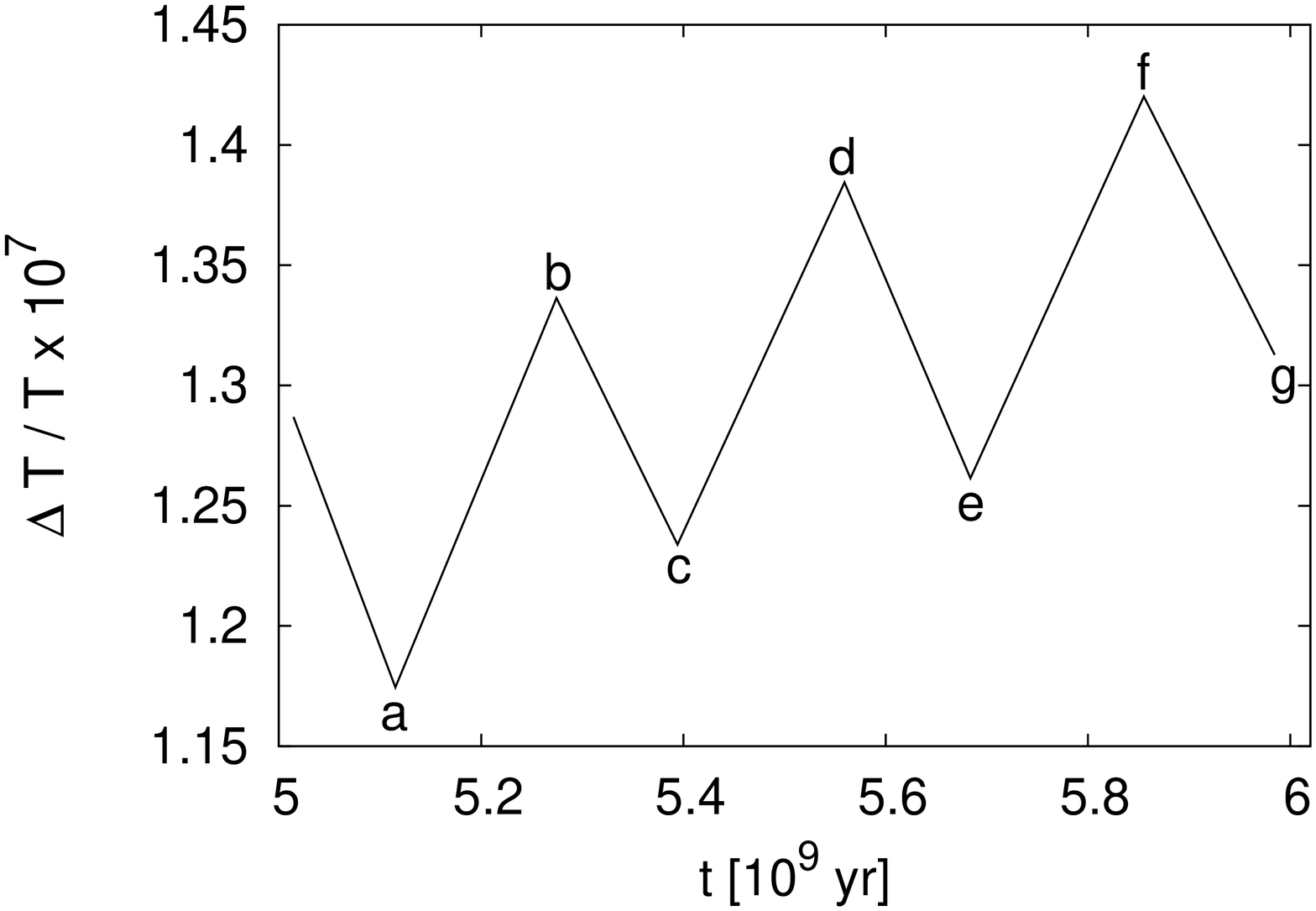}    
\caption{A small part of the light propagation in model 3. The left panel shows
the density variation that light `feels' as it propagates. The black thin dotted
line shows the density in the background model. The right panel presents the
temperature fluctuations as measured by an observer situated outside the
structure in the homogeneous FLRW region. The letters in the left and right 
panels label corresponding points along the light path.}
\label{fig4} 
\end{figure}

To estimate the temperature fluctuations induced by the light propagation
effects, it is assumed that initial temperature distribution is uniform,
$(\Delta T/T)_e = 0$. Then temperature fluctuations are calculated using
(\ref{7.3}), and they are plotted against time of propagation in Fig.\
\ref{fig3}. As seen, the final values are small, of amplitude $\Delta T/ T
\approx 10^{-7}$ (model 3), $\Delta T/ T \approx 10^{-6}$ (models 1 and 2), and
$\Delta T/ T \approx 10^{-5}$ (models 4 and 5). A detailed analysis of how
inhomogeneities induce temperature fluctuations is presented in Fig.\
\ref{fig4} (for clarity, only a small fraction of the time is presented). The
left panel of Fig.\ \ref{fig4} shows the density of regions through which the
light propagates in model 3. The right panel presents the temperature
fluctuations as measured by an observer situated at the junction point where the model is that of
Friedmann. Letters correspond to each inhomogeneous patch (left panel) and
temperature fluctuations caused by them (right panel). Clearly, underdense
regions induce negative temperature fluctuations, overdense regions induce
positive fluctuations.

\begin{figure}
\includegraphics[scale=0.55]{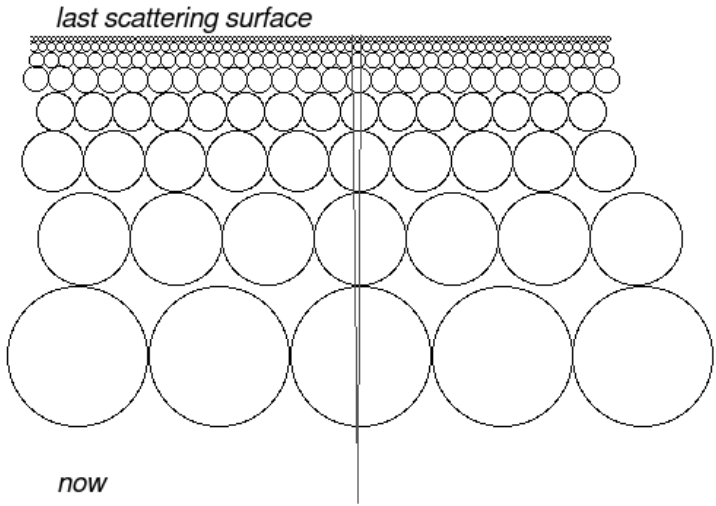}
\caption{The schematic representation of the Swiss Cheese model.
When two photons are propagating along a similar path the final
temperature fluctuations are similar. If paths are different,
then the final temperature fluctuations are also different and hence not correlated.}
\label{fig5}
\end{figure}

\begin{figure}
\includegraphics[scale=0.33]{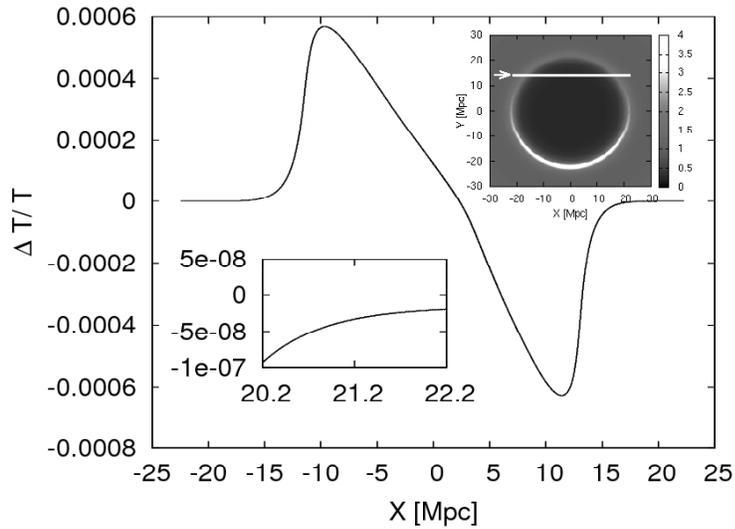}
\caption{Temperature fluctuation amplitude, as measured by observers at different locations in region E, along the path of a light ray. The ray path is shown as the bright line in the upper inset.}
\label{fig6}
\end{figure}

Apart from estimating
the amplitude of the Rees--Sciama effect, it is also important to estimate the
angular scale which is the most affected by this effect. Without going into any
complicated analysis, we can estimate the angular scale by employing the
following approximation: the correlation between two distant points on the sky
is zero -- photons which were propagating along two
distant paths have the temperature fluctuations uncorrelated. Only when the light paths
are near to each other
are the temperature fluctuations correlated. Thus the simplest estimation of the
angular scale of the Rees--Sciama effect, as seen from the schematic Fig.\
\ref{fig5}, is the angular size of the Szekeres patch at the last scattering
instant. For the models studied in this section, such approximations lead to an
angular scale of $\vartheta \approx 0.21^{\circ}$, or alternatively $\ell \approx
850$. If the photons are propagating along
neighbouring paths for only half of the age of the Universe (in such a case, as seen from
Fig.\ \ref{fig6}, the final temperature fluctuations are two times smaller),
then the angular scale is similar,  $\vartheta \approx 0.24^{\circ}$ ($\ell \approx
750$). Thus, the Rees--Sciama effect of amplitude $\sim 10^{-6}$ contributes to
the CMB temperature fluctuations on the angular scale $\vartheta < 0.25^{\circ}$
($\ell > 700$). This angular scale corresponds to the angular scale at which the
third peak of the CMB angular power spectrum is observed. At this scale the
measured rms temperature fluctuations are of amplitude $\approx 5 \times
10^{-5}$. This is still several times higher than the results obtained within
models 4 and 5. In the case of models 1--3 the measured value is more than one
order of magnitude larger than the model estimates.

\subsection{The role of local structures}\label{loktemp}

So far, it has been assumed that each inhomogeneous structure is compensated
(i.e.\ each Szekeres region was matched with the FLRW background), and that
measurements are carried out away from the inhomogeneities, i.e.\ where the
universe is homogeneous. However in the real Universe there is no place where the cosmic structures in the observer's vicinity are fully compensated and therefore the Universe 
 should not be treated locally as homogeneous. Since all
measurements are always local, let us consider what happens if temperature fluctuations are measured in an uncompensated region. Figure \ref{fig6}
presents the temperature fluctuations measured by an observer situated at
different places within region E. 
These results  are obtained under the assumption that light from last scattering is
propagating through homogeneous regions, and currently reaches an observer
in an inhomogeneous structure (region E). The light enters and
propagates along the bright line shown in the upper right inset of 
Fig.\ \ref{fig6}. 
The above results show that local structures
can significantly contribute to the CMB temperature fluctuations. This indicates
that care must be exercised when extracting information from the CMB
observations. Although it is highly unlikely that the signal caused by the local
structures have a signature of acoustic oscillations we should be aware that
local structures can have some visible impact on observations. 
Thus, it is important to test if local structures can 
cause the observed correlations of the alignment of dipole, quadrupole and octopole axes of angular power spectrum of the CMB temperature fluctuations (cf. \cite{SSHC,LaM1,LaM2,Vale,RRS,RaSch}) or their low amplitude (cf. \cite{PP90,LP96,SC99}).

\section{Conclusions}

The analysis presented in this paper has aimed to examine the influence of the light
 propagation effects on the temperature fluctuations of the CMB. 
The results  indicate that the Rees-Sciama effect caused by the propagation of light
through  inhomogeneous but compensated structures do not significantly affect the CMB temperature fluctuations. Some would say that this is
an obvious result since similar conclusion is reached
when using perturbative methods. However, as it was argued in the 
Sec. \ref{intro}, within real cosmic structures 
the density contrast and the Weyl curvature are significantly large.
Thus, the application of the perturbation methods cannot be justified.
It was shown, in this paper, that even in such cases light propagation
effects are  small\footnote{However, in this paper all inhomogeneities
are at the current instant of diameters of $\sim 50-60$ Mpc. Larger inhomogeneities,
of diameters $\sim 600$ Mpc, can have more significant impact, cf. \cite{IS06,IS07,MN08}.}.
It has also been shown that the  Rees--Sciama effect of amplitude $\sim 10^{-6}$ contributes 
to the CMB temperature fluctuations on the angular scale $\vartheta < 0.25^{\circ}$ ($\ell > 700$).

However, if the structures are not compensated or the measurements are carried out inside
inhomogeneous non-compensated structures, the amplitude of measured temperature fluctuations can be slightly higher. 
Since in reality we cannot separate ourselves from the surroundings and
say that all local structures at our positing are ``compensated'',
thus the local cosmic structures must be taken into account when analysing 
CMB observations. Especially, it is possible that the local structures can have some impact on low multipoles anomalies of the angular CMB power spectrum.

\begin{acknowledgements}
The Peter and Patricia Gruber Foundation and the International Astronomical Union 
are gratefully acknowledged for support.
I would like to thank Charles Hellaby, Andrzej Krasi\'nski, Paulina Wojciechowska and
the referees for useful discussions, comments and suggestions.
This research has been partly supported by Polish Ministry of Science and Higher Education under grant N203 018 31/2873, allocated for the period 2006-2007, and by the Polish Astroparticle Network (621/E-78/SN-0068/2007).
\end{acknowledgements}

\appendix

\setcounter{section}{0}

\section{The redshift formula for axial geodesics}\label{app1}

To find a simplified redshift formula for the axial geodesics first we
have to find $k^0$ in an affine parametrisation and then use the relation
(\ref{rdfdef}). Since we assume that $k^2 = k^3 = 0$ then let us choose
\begin{equation}
k^1 = -1,~~~~~~~~k^0 = \frac{\Phi,_r - \Phi E,_r/E}{\sqrt{1 - k}}.
\label{ape1}
\end{equation}
The above parametrisation is not affine, thus the parallel transport does not
preserve the tangent vector. Therefore $k^{\alpha}$, after being parallely
transported, becomes $\lambda k^{\alpha}$ (where $\lambda$ is a scalar
coefficient -- a function of the parameter $s$ along the geodesic). In this case
the geodesic equations are of form \cite{PK06}
\begin{equation}
k^{\alpha};_{\beta} k^{\beta} = - \frac{1}{\lambda} \frac{{\rm d} \lambda}{{\rm
d} s} k^{\alpha}, \label{ape2}
\end{equation}
and reduce to
\begin{eqnarray}
&& - \frac{1}{\lambda} \frac{{\rm d} \lambda}{{\rm d} s}  
= - \frac{ \Phi,_{rr} - \Phi,_r E,_r/E -
\Phi E,_{rr}/E +
\Phi (E,_r/E)^2}{\Phi,_r - \Phi E,_r/E}
\nonumber \\
&& +  2 \frac{{\Phi},_{tr}  - \Phi,_t E,_r/E}{\sqrt{1 - k}} 
- \frac{k,_r}{2(1 - k)}.
\label{ape3}
\end{eqnarray}
All the quantities above are evaluated on the null geodesic.
In order to better depict which quantity is evaluated on the null geodesic
a symbol $~\hat{}~$ is be used.
Since on the geodesic
$t$ and $r$ are connected with each other via relation (\ref{snge}), we have
\begin{eqnarray}
&& \Phi_n,_r = (\Phi,_t )_n \frac{{\rm d} t}{{\rm d} r} + (\Phi,_r)_n
\nonumber \\
&& (\Phi,_r)_n,_r = (\Phi,_{tr} )_n \frac{{\rm d} t}{{\rm d} r} +
(\Phi,_{rr})_n, \label{ndrv}
\end{eqnarray}
where the subscript $n$ referees to quantities measured on the geodesic.

The second term in equation (\ref{ape3}) looks like a logarithmic derivative.
However because of (\ref{ndrv}) we have
\begin{eqnarray}
&& \frac{ {\rm d} \ln \left[ (\Phi,_r)_n - \Phi_n (E,_r/E)_n
\right] }{ {\rm d} r} = \nonumber \\
&& \frac{ (\Phi,_{r})_n,_r - \Phi_n,_r (E,_r/E)_n - \Phi_n ({E},_{rr}/E)_n + \Phi_n (E,_{r}/E)^2_n}{ (\Phi,_r)_n -
\Phi_n (E,_r/E)_n} = \nonumber \\
&& \frac{ (\Phi,_{rr})_n - (\Phi,_r)_n (E,_r/E)_n - \Phi_n ({E},_{rr}/E)_n + \Phi_n (E,_{r}/E)^2_n}{
(\Phi,_r)_n - \Phi_n (E,_r/E)_n} \nonumber \\
&& + \frac{(\Phi,_{tr})_n - (\Phi,_t)_n (E,_r/E)_n}{ (\Phi,_r)_n -
\Phi_n (E,_r/E)_n} \frac{{\rm d} t}{{\rm d} r}.
\label{ngeo00b}
\end{eqnarray}
Using the above relation we can integrate equation (\ref{ape3})
\begin{equation}
\lambda = C \frac{\sqrt{1 - k}}{(\Phi,_r)_n - \Phi_n (E,_r/{E})_n}~\exp \left( \int {\rm d} r \frac{ (\Phi,_{tr})_n - (\Phi,_t)_n ({E},_r/E)_n}{\sqrt{1 - k_n}}. \right) \label{ngeo01b}
\end{equation}
Now we can easily find that $k^{\alpha}$ in the affine parametrisation is given
by ${\tilde k}^{\alpha} = (\lambda/C) k^{\alpha}$. Using 
(\ref{rdfdef}) we obtain
\begin{equation}
\ln (1+z) =  \pm \int\limits_{r_e}^{r_o} {\rm d} r \frac{ (\Phi,_{tr})_n -
(\Phi,_t)_n (E,_r/E)_n}{\sqrt{1 - k_n}},
\label{zintr}
\end{equation}
where $+$ is for ${r_e} < {r_o}$ and
$-$ for ${r_e} > {r_o}$. Alternatively the redshift
can be found by integration over time:
\begin{equation}
\ln (1+z) =  \int\limits_{t_e}^{t_o} {\rm d} t
\frac{ (\Phi,_{tr})_n -
(\Phi,_t)_n (E,_r/E)_n}{(\Phi,_r)_n - \Phi_n (E_r/E)_n}.
\label{zintt}
\end{equation}

\section{Model specification and evaluation}\label{app2}

In order to define the Szekeres model five functions of radial coordinate needed
to be specified. In this paper all models will be defined by the following set of functions: $k,~M,~S,~P,$ and
$Q$.

The algorithm used in the calculations can be defined as follows:

\begin{enumerate}
 \item
The radial coordinate is chosen to be the areal radius at the last scattering instant
$r' = \Phi(r, t_{LS})$.
However, for clarity in further use, the prim is omitted
and the new radial coordinate will be referred to as $r$.

\item
The chosen background model is the $\Lambda$CDM model, i.e. a flat FLRW 
model with $\Lambda \ne 0$. The background density at the current instant is then
given by

 \begin{equation}
   \rho_b = \Omega_m \times \rho_{cr} = 0.27 \times \frac{3H_0^2}{8 \pi G}.
   \label{4.51}
 \end{equation}
where the Hubble constant is $H_0 =72$ km s$^{-1}$ Mpc$^{-1}$.  The cosmological
constant,  $\Lambda$, corresponds to $\Omega_{\Lambda} = 0.73$, where
$\Omega_{\Lambda} = (1/3)  ( c^2 \Lambda/H_0^2)$.

\item
The initial time, $t_i$, is chosen to be the time of last scattering,
and is calculated from the following formula for a background FLRW universe \cite{P80}

\begin{equation}
t(z) =  \frac{1}{H_0} \int\limits_{z}^{\infty} \frac{{\rm d}
\tilde{z}}{(1+\tilde{z}) \sqrt{ \mathcal{D}(\tilde{z})} }, \label{4.27}
\end{equation}
where:
\begin{equation}
\mathcal{D} (z) = \Omega_{m} (1+z)^3 + \Omega_K
(1+z)^2 + \Omega_{\Lambda},
\end{equation}
where $\Omega_K = 1 - \Omega_{m} - \Omega_{\Lambda}$.
For the lower limit of integration, $z = 1089$ was used as the redshift at last scattering.

\item
Six different Szekeres regions are considered in this paper.
Let us denote them as regions A, B, C, D, E and F.
The functions  $M,~k,~Q,~P$ and $S$ in these regions are defined as follows

\begin{itemize}
\item regions A and B
\[  M = M_b + \left\{ \begin{array}{ll}
M_1 r^3 & {\rm ~for~} r \leqslant 0.5a, \\
M_2 \exp \left[ - 12 \left( \frac{ r - a}{a} \right)^2 \right] & {\rm ~for~}
0.5a \leqslant  r \leqslant 1.5a \\
M_1 (2a - r)^3 & {\rm ~for~} 1.5a \leqslant  r \leqslant  2a,  \\
0 &  {\rm ~for~} r \geqslant 2a,
\end{array} \right. \]
\noindent where $M_b$ is the mass in the corresponding volume of the homogeneous
universe, i.e.\ $M_b = (4 \pi G /3c^2) \rho_{LS} r^3$, $\rho_{LS} = \rho_b (1+z_{LS})^3$, $M_1 = 8 M_2 a^{-3} {\rm e}^{-3/2}$, $M_2$ is equal to $-0.3$
kpc and $0.2$ kpc for region A and B respectively, and $a=12$ kpc.
\[ k = - \frac{1}{2} \times  \left\{ \begin{array}{ll}
k_1 r^2 & {\rm ~for~} r \leqslant 0.5 b, \\
k_2 \exp \left[ - 4 \left( \frac{ r - b}{b} \right)^2 \right] & {\rm ~for~} 0.5
b \leqslant  r \leqslant 1.5 b \\
k_1 (2b - r)^2 & {\rm ~for~} 1.5 b \leqslant  r \leqslant  2 b,  \\
0 &  {\rm ~for~} r \geqslant 2b,
\end{array} \right. \]
\noindent
where $k_1 = 4 k_2 a^{-2} {\rm e}^{-1}$, $k_2$
is equal to $-5.15 \times 10^{-6}$ and $3.5 \times 10^{-6}$
for regions A and B respectively, and $b=10.9$ kpc.
\[
S = 1, \quad P = 0, \quad  Q = Q_1 \ln (1+ Q_2 r) \times \exp (- Q_3 r),
\]
where, for regions A and B respectively, $Q_1$ equals $-0.72$  and
$-1.45$, $Q_2$ equals $1$ kpc$^{-1}$ and $0.4$ kpc$^{-1}$, and $Q_3$
equals $0.01$ kpc$^{-1}$ and $0.005$ kpc$^{-1}$. With these definitions the mass
distribution and the curvature are the same as in Friedmann models, for $r>24$
kpc.

\item Regions C$_1$ and C$_2$

In region C the functions $M$ and $k$ are the same as in region A. The only
difference is in the form of functions $S$, $P$, and $Q$ which are as follows
\[
S = {\rm e}^{\alpha r}, \quad  P = 0, \quad  Q = 0,\] where $\alpha$ equal to
$-0.0255$ kpc$^{-1}$ and $+0.0255$ kpc$^{-1}$ for regions C$_1$ and C$_2$
respectively. Region C$_1$ is the mirror image of C$_2$,
where the $Z=0$ surface is the symmetry plane [$Z = \Phi \cos \vartheta$ and
$\vartheta$ is defined by the stereographic projection (\ref{stp})].
The reason for employing two mirror-similar regions is that in the
coordinates used here, the axial geodesics can only be studied for propagation
along the $Z<0$ direction, in which $\vartheta = - \pi$. Along the $Z>0$
direction we have $\vartheta = 0$, which corresponds to a point at infinity in
the stereographic projection.
This problem is overcome by matching C$_1$ with C$_2$ along the surface of
$Z=0$. When calculating propagation toward the origin model C$_1$ is employed,
and when calculating propagation away from the origin model C$_2$ is employed.
In both models light propagates along the $Z<0$ axis.

\item Regions D$_1$ and D$_2$

In region D the functions $M$ and $k$ are the same as in region B. The only
difference is in the form of the functions $S$, $P$, and $Q$ which are of the
following form:

\[
S = r^{\alpha}, \quad P = 0, \quad  Q = 0,
\]
where $\alpha$ equal to $-0.97$ and $+ 0.97$ for regions D$_1$ and D$_2$
respectively. As above, region D comes from matching regions D$_1$ and D$_2$
along the $Z=0$ surface.

\item Regions E and F

\[
   M = M_b + \left\{ \begin{array}{ll}
      M_1 r^3 & {\rm ~for~} r \leqslant 0.5 a, \\
M_2 \exp \left[ - 6 \left( \frac{ r - a}{a} \right)^2 \right] & {\rm ~for~} r
\geqslant 0.5 a,
      \end{array} \right. \]
\[
   k  = \left\{ \begin{array}{ll}
      k_1 r^2 & {\rm ~for~} r \leqslant 0.5 b,\\
k_2 \exp \left[- \left( \frac{ r - b}{0.5 b} \right)^2 \right] & {\rm ~for~} r
\geqslant 0.5 b,
      \end{array} \right. \]
\[
   S  = 1,~~~P = 0,~~~Q = Q_1 -0.22 \ln (1+ Q_2 r) \times \exp (- Q_3 r).
 \]
where $M_1 = 8 a^{-3} M_2 {\rm e}^{-1.5}$, $k_1 = 4 a^{-2} k_2 {\rm e}^{-1}$.
For region E,  $M_2 = -0.75$ kpc,
 $a = 15.23$ kpc, $k_2 = -1.00173 \times
10^{-5}$, $b = 12.95$ kpc, $Q_1 = -0.22$, $Q_2 = 1$ kpc$^{-1}$, $Q_3 = 0.1$
kpc$^{-1}$. For region F,
$M_2 = 0.9$ kpc, $a = 23.76$ kpc, $k_2 = 7
\times 10^{-6}$, $b = 19.1$ kpc, $Q_1 = -1.4$, $Q_2 = 0.4$ kpc$^{-1}$, $Q_3 =
0.005$ kpc$^{-1}$.

\end{itemize}

\item
Light propagation was calculated by solving eqs. 
(\ref{lp0}) -- (\ref{lp3}) (models 1 and 3)
and (\ref{snge}) (models 2, 4 and 5)
simultaneously with the evolution equation (\ref{vel}).
At each step the null condition, $k_{\alpha}k^{\alpha} = 0$ was used to
to test the precision of calculations. 
All equations were solved using the fourth order Runge--Kutta method.

\item
The temperature fluctuations were calculated from eq. (\ref{dtpt}). 
The redshift was calculated using relation 
(\ref{rdfdef}) (models 1 and 3)
and (\ref{srf}) (models 2, 4 and 5).
The mean redshift $\bar{z}$ was calculated using the ${\rm \Lambda CDM}$ model.

\end{enumerate}

\end{document}